\documentclass[12pt,preprint]{revtex4}
\usepackage{graphicx} 
\usepackage{amsmath}
\usepackage{amsfonts,amssymb}
\usepackage[matrix,arrow,curve,frame,poly,arc]{xy}
\usepackage[english]{babel}

\def\be{\begin{equation}}
\def\ee{\end{equation}}
\def\ba{\begin{eqnarray}}
\def\ea{\end{eqnarray}}

\begin{document}
\bibliographystyle{plainnat}

\title{Exact traveling wave solutions of 1D parabolic-parabolic models of chemotaxis}

\author{Maria Shubina }

\email{yurova-m@rambler.ru}

\affiliation{Skobeltsyn Institute of Nuclear Physics\\Lomonosov Moscow State University
\\ Leninskie gory, GSP-1, Moscow 119991, Russian Federation}


\begin{abstract}

In this paper we consider three different 1D parabolic-parabolic systems of chemotaxis. For these systems we obtain the exact analytical solutions in terms of traveling wave variables. 

\end{abstract}

\keywords{parabolic-parabolic system, exact solution, soliton solution, Patlak-Keller-Segel model}

\maketitle

\section{Introduction}
\label{intro}

In this paper we consider a number of different systems of nonlinear partial differential equations, which describe a directed cells (bacteria or other organisms) movement up or down a chemical concentration gradient (chemotaxis). The aim of this paper is to obtain exact analytical solutions of these models. For 1D parabolic-parabolic systems under consideration we present these solutions in explicit form in terms of traveling wave variables. Of course, not all of the solutions obtained can have appropriate biological interpretation since the biological functions must be nonnegative in all domain of definition. However some of these solutions are positive and bounded and their analysis requires further investigation. 

Chemotaxis plays an important role in many biological and medical fields such as embryogenesis, immunology, cancer growth. The macroscopic classical model of chemotaxis was proposed by Patlak in 1953 \cite{P} and by Keller and Segel in the 1970s \cite{KS1}-\cite{KS3}. This model describes the space-time evolution of a cells density $ u(t,\overrightarrow{r}) $ and a concentration of a chemical substance $ v(t,\overrightarrow{r}) $. The general form of this model is:
\be \left\{
\begin{aligned}
u_{t}-\nabla (\delta_{1} \nabla u - \eta_{1} u \nabla \phi (v)) & =0 \\ 
v_{t}-\delta_{2} \nabla^{2} v - f(u, v) & =0, \nonumber
\end{aligned}
\right.
\ee
where $ \delta_{1} > 0 $ and $ \delta_{2} \geq 0  $ are cells and chemical substance diffusion coefficients respectively, $ \eta_{1} $ is a chemotaxis coefficient; when $ \eta_{1} > 0 $ this is an attractive chemotaxis ("positive taxis"), and when $ \eta_{1} < 0 $ this is a repulsive ("negative") one \cite{Ni}, \cite{Li&Wang}. The function $ \phi (v) $ is the chemosensitivity function and $ f(u, v) $ characterizes the chemical growth and degradation. These functions are taken in different forms that corresponds to some variations of the original Keller--Segel model. We follow the reviews of T. Hillen and K. Painter \cite{Hillen&Painter} and of  Z.-A. Wang \cite{Wang} and consider models presented therein. 

This paper is concerned with one-dimensional simplified models when the coefficients $ \delta_{1} $, $ \delta_{2} $ and $ \eta_{1} $ are positive constants, $ x \in \Re, t \geq 0 $, $ u=u(x,t) $, $ v=v(x,t) $.

\section{Signal-dependent sensitivity model}
\label{sec:1}

Let us start with a model that allows nonnegative bounded solutions that may be of interest from a biological point of view. Now consider the "logistic" model, one of versions of signal-dependent sensitivity model \cite{Hillen&Painter} with the chemosensitivity function $ \phi (v) = (1 + \textit{b})   \ln(v + \textit{b}) $, $  \textit{b} = const $ and $ f(u, v)= \tilde{\sigma} u - \tilde{\beta} v  $. In the review \cite{H1} one can see a mathematical analysis of this model. When $ \textit{b} = 0 $ and $ \tilde{\beta} = 0 $ the existence of traveling waves were established in \cite{N&I}, \cite   {E&F&N}. The replacement $ t \rightarrow  \delta_{1} t $, $ u \rightarrow \sigma \dfrac{\tilde{\sigma} }{\delta_{1}} u $ gives $\delta_{1} = 1 $, $ \alpha = \dfrac{\delta_{2}}{\delta_{1}} $, $ \beta = \dfrac{\tilde{\beta}}{\delta_{1}} $, $ \sigma = \pm 1  $. We also set $ \eta = \dfrac{\eta_{1} (1 + \textit{b})}{\delta_{1}} $, $ 1 + \textit{b} > 0 $, as well as $ \phi (v) =  \ln|v + \textit{b}| $. It should be noted that a sign of $ \sigma $ may affect on mathematical properties of the system. So, $ \sigma = 1 $ corresponds to an increase of a chemical substance, proportional to cells density, whereas $ \sigma= - 1 $ corresponds to its decrease. And as we shall see later, various solutions correspond to these two cases. 

After above replacements the model reads:
\be 
\label{eq:1} 
\left\{
\begin{aligned}
u_{t}- u_{xx} + \eta (u \frac{v_{x}}{v + \textit{b}})_{x} & =0 \\ 
v_{t}-\alpha v_{xx} - \sigma u  + \beta v & =0.\\ 
\end{aligned}
\right.
\ee
If we introduce the function $ \upsilon = v + \textit{b} $, in terms of traveling wave variable $ y = x - ct $, $ c = const $ this system has the form:
\be 
\left\{
\begin{aligned}
u_{y} + c u -  \eta u (\ln ( \upsilon ))_{y}+ \lambda & =0 \\ 
\alpha \upsilon_{yy} + c \upsilon_{y} - \beta \upsilon + \beta \textit{b} + \sigma u & =0,\\ 
\end{aligned}
\right. 
\tag{\ref{eq:1}$*$} 
\ee
where $ u=u(y) $, $ \upsilon = \upsilon (y) $ and $ \lambda $ is an integration constant.

In this paper we will consider the case of $ \lambda = 0 $. Then the first equation in ($1*$) gives
\be
u = C_{u} e^{-cy} \upsilon^{\eta},
\ee
$ C_{u} $ is a constant and we will examine the following equation for $ \upsilon $:
\be
\alpha \upsilon_{yy} + c \upsilon_{y} - \beta \upsilon + \beta \textit{b} + \sigma C_{u} e^{-cy} \upsilon^{\eta} = 0.
\ee 
Since $ \eta $ is a positive constant we consider two cases: $ \eta = 1 $ and (3) is linear nonhomogeneous equation, and $ \eta \neq 1 $.

\subsection{$\eta = 1$}

Let us begin with $ \eta = 1 $. We introduce the new variable $ z $ and the new function $ w $:
\ba
z & = & \left( \dfrac{4 \sigma C_{u} }{\alpha c^{2 }} \right) ^{\frac{1}{2}}\,\,e^{- \frac{c y}{2 }}\\
\nonumber
w & = & \left( \dfrac{4 \sigma C_{u} }{\alpha c^{2 }} \right) ^{\frac{\alpha - 2}{4 \alpha}}\,\, \upsilon \, e^{\frac{c y}{2 \alpha }}
\ea
and equation (3) becomes:
\be
z^{2} w_{zz} +z w_{z} + w ( z^{2} - \nu^{2})  = \Lambda z^{- \frac{1}{\alpha}},
\ee
where $ \nu^{2} = \dfrac{1}{\alpha^{2}} (1 + \dfrac{4 \alpha \beta }{c^{2}} ) $, $ \Lambda = - \dfrac{4 \beta \textit{b} }{\alpha c^{2}}\left( \dfrac{ 4 \sigma C_{u}}{\alpha c^{2}} \right)^{\frac{1}{4}} $. Equation (5) is the Lommel differential equation \cite{Bateman&Erdelyi}, \cite{Watson} with $ \mu = -1 - \frac{1}{\alpha} $. For $ \sigma C_{u} > 0 $ its general solution has the form:
\be
w(z) = C_{J} J_{\nu} (z) + C_{Y} Y_{\nu} (z) + \Lambda S_{\mu, \nu} (z),
\ee
where $ C_{J} $, $ C_{Y} $ are constants, $ J_{\nu} (z) $ and $ Y_{\nu} (z) $ are Bessel functions and 
\ba
S_{\mu, \nu} (z) & = & s_{\mu, \nu} (z) + 2^{\mu - 1} \Gamma ( \dfrac{\mu - \nu + 1}{2} ) \, \Gamma ( \dfrac{\mu + \nu + 1}{2}) \, \left[ \sin (\frac{\pi}{2}(\mu - \nu))J_{\nu} (z) - \cos (\frac{\pi}{2}(\mu - \nu))Y_{\nu} (z) \right], \nonumber 
\\
s_{\mu, \nu} (z) & = & \dfrac{z^{\mu + 1}}{[(\mu + 1)^{2} - \nu^{2}]}\,\, _{1}F_{2}( 1; \dfrac{\mu - \nu + 3}{2}, \dfrac{\mu + \nu + 3}{2}; - \dfrac{ z^{2}}{4} ) 
\ea
are Lommel functions, $ _{1}F_{2} $ is generalized hypergeometric function \cite{Bateman&Erdelyi}, \cite{Watson}. Further, substituting of the initial variable $ y $ and the function $ v $ (see (4)) into (6) we obtain a formal solution.

\subsubsection{$\textit{b}  = 0$}

We first consider the case $ \textit{b}  = 0$. Then $ \upsilon = v \geq 0 $ and $ C_{u} > 0 $. Equation (5) becomes homogeneous and for $ \sigma = 1 $ its general solution is 
\be
w(z) = C_{J} J_{\nu} (z) + C_{Y} Y_{\nu} (z). 
\ee
However one can check that the function $ u=u(y) $ diverges as $ c y \rightarrow - \infty $ for all $ \nu $. 

Consider now $ \sigma = - 1 $. For $ v(y) $ be real let $ \alpha = 2 $. Then (5) becomes the modified Bessel equation; the analysis of solutions behavior at $ \pm \infty $ leads to suitable solutions for $ v(y) $ and $ u(y)$:
\ba
v(y) & = & e^{-\frac{cy}{4}}\,K_{\nu}(\sqrt{\frac{2C_{u}}{c^{2}}}\,\,e^{-\frac{cy}{2}})\\
\nonumber
u(y) & = & C_{u} e^{-\frac{5cy}{4}}\,K_{\nu}(\sqrt{\frac{2C_{u}}{c^{2}}}\,\,e^{-\frac{cy}{2}})
\ea
with restrictions $ \nu \leq \dfrac{1}{2} $ and $ \beta \leq 0 $. So on can see that $ v(y) \rightarrow 0 $ as $ cy \rightarrow - \infty $ for all $ \nu \leq \dfrac{1}{2} $; $ v(y) \rightarrow 0 $ for $ \nu < \dfrac{1}{2} $ and $ v(y) \rightarrow \sqrt[4]{\dfrac{\pi^{2} c^{2}}{8 C_{u}}} $ for $ \nu = \dfrac{1}{2} $ as $ cy \rightarrow \infty $ and $ u(y)\rightarrow 0 $ as $ y \rightarrow \pm \infty $ for all $ \nu \leq \dfrac{1}{2}  $. The curves of these functions are presented in Fig.1--Fig.2. Thus, the solution obtained may be considered as biologically appropriated one and this requires further investigation.

\begin{figure}[h!]
\begin{minipage}{0.49\linewidth}
\center{\includegraphics[width=1\linewidth]{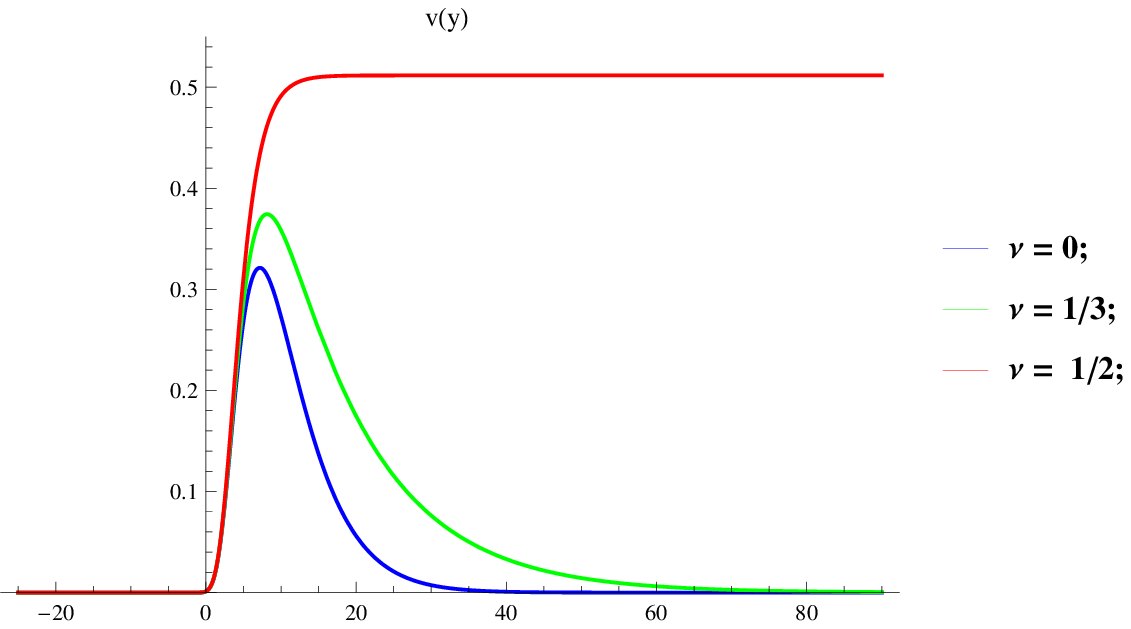} \\ Fig.1:  $ v(y) $; $c=1$; $C_{u}=18$; }
\end{minipage}
\hfill
\begin{minipage}{0.49\linewidth}
\center{\includegraphics[width=1\linewidth]{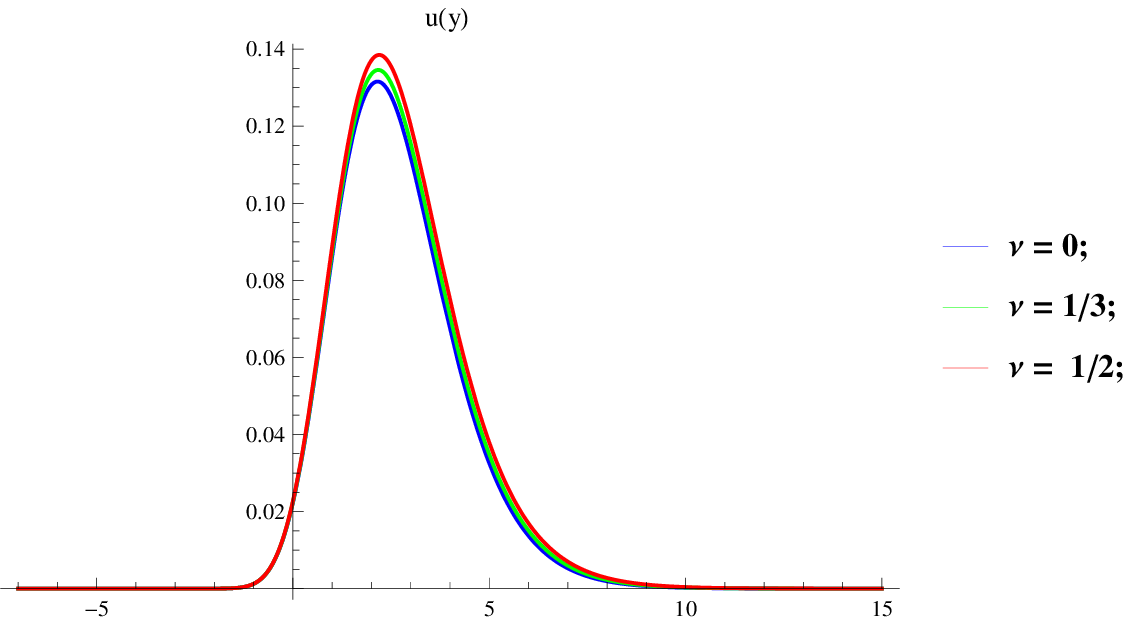} \\ Fig.2:  $ u(y) $; $c=1$; $C_{u}=18$;}
\end{minipage}
\end{figure}

\subsubsection{$\textit{b}  > 0$}

Let us return to equation (5) with $ \Lambda \neq 0 $. The analysis of solutions asymptotic forms at $ \pm \infty $ \cite{Bateman&Erdelyi}, \cite{Watson} gives the following expressions for $ v(y) $ and $ u(y)$:
\ba
v(y) + \textit{b} & = & - \dfrac{4 \beta \textit{b} }{\alpha c^{2}}\left( \dfrac{ 4  \sigma C_{u}}{\alpha c^{2}} \right)^{\frac{1}{2 \alpha}}\,\,e^{- \frac{c y}{2 \alpha }} \,\,S_{\mu, \nu} (\sqrt{ \dfrac{4  \sigma C_{u} }{\alpha c^{2 }} }\,\,e^{- \frac{c y}{2 }})\\
\nonumber
u(y) & = & - C_{u}\, \dfrac{4 \beta \textit{b} }{\alpha c^{2}}\left( \dfrac{ 4  \sigma C_{u}}{\alpha c^{2}} \right)^{\frac{1}{2 \alpha}}\,\,e^{- cy ( 1 +  \frac{1}{2 \alpha })} \,\,S_{\mu, \nu} (\sqrt{ \dfrac{4  \sigma C_{u} }{\alpha c^{2 }} }\,\,e^{- \frac{c y}{2 }})
\ea
with $ \sigma C_{u} > 0 $ and for $ \nu < \dfrac{1}{\alpha} $.  The latter condition leads to the requirement $ -\dfrac{c^{2}}{4\alpha} \leq \beta < 0 $. The $ v(y) \rightarrow -\textit{b} $ and $ u(y)\rightarrow -\dfrac{\beta \textit{b}}{\sigma} $ as $ cy \rightarrow - \infty $ and $ v(y) \rightarrow 0 $, $ u(y)\rightarrow 0 $ as $ cy \rightarrow  \infty $. Thus, one can see that for $ \textit{b} > 0$, $ \sigma = 1 $ and $ C_{u} > 0 $ $ u(y) \geq 0 $ is satisfied but $ v(y) < 0 $. These functions are presented in Fig.3--Fig.4. It should be noted that $ \nu \neq \dfrac{1}{\alpha}  $, or $ \beta \neq 0 $ because of pole in $ \Gamma $ - function.
\begin{figure}[h!]
\begin{minipage}{0.49\linewidth}
\center{\includegraphics[width=1\linewidth]{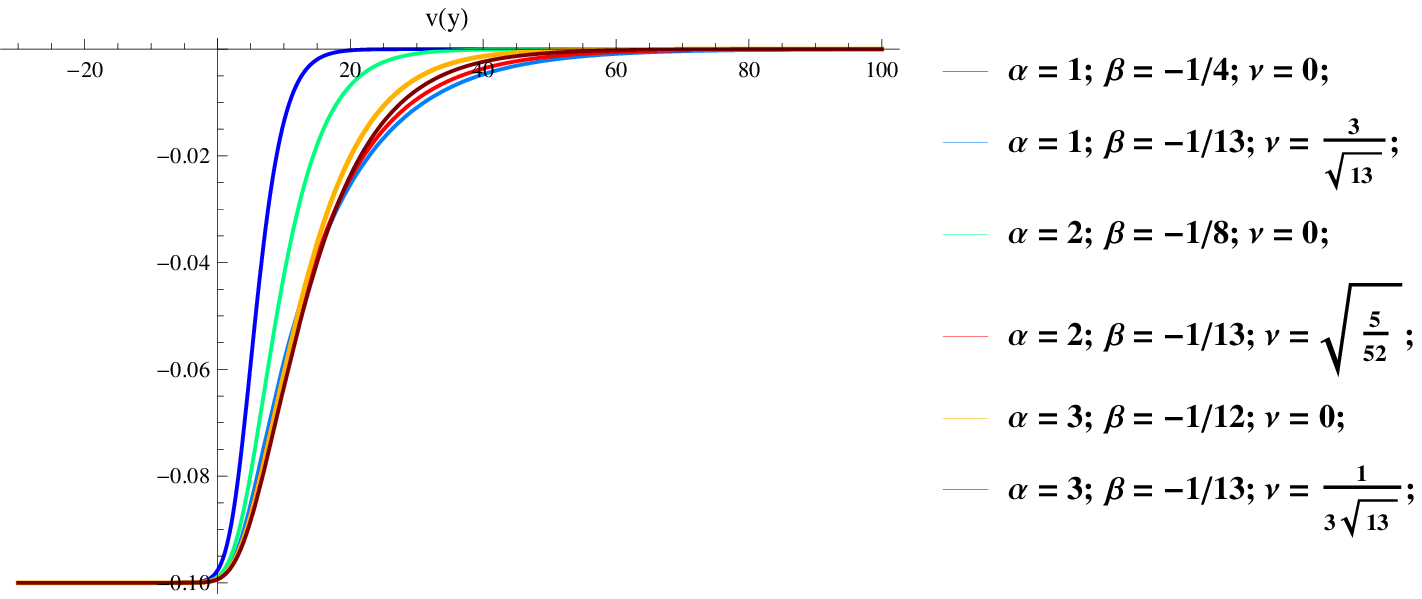} \\ Fig.3:  $ v(y) $; $c=1$; $C_{u}=9$; $ \sigma = 1 $; $ \textit{b}= 0.1 $}
\end{minipage}
\hfill
\begin{minipage}{0.49\linewidth}
\center{\includegraphics[width=1\linewidth]{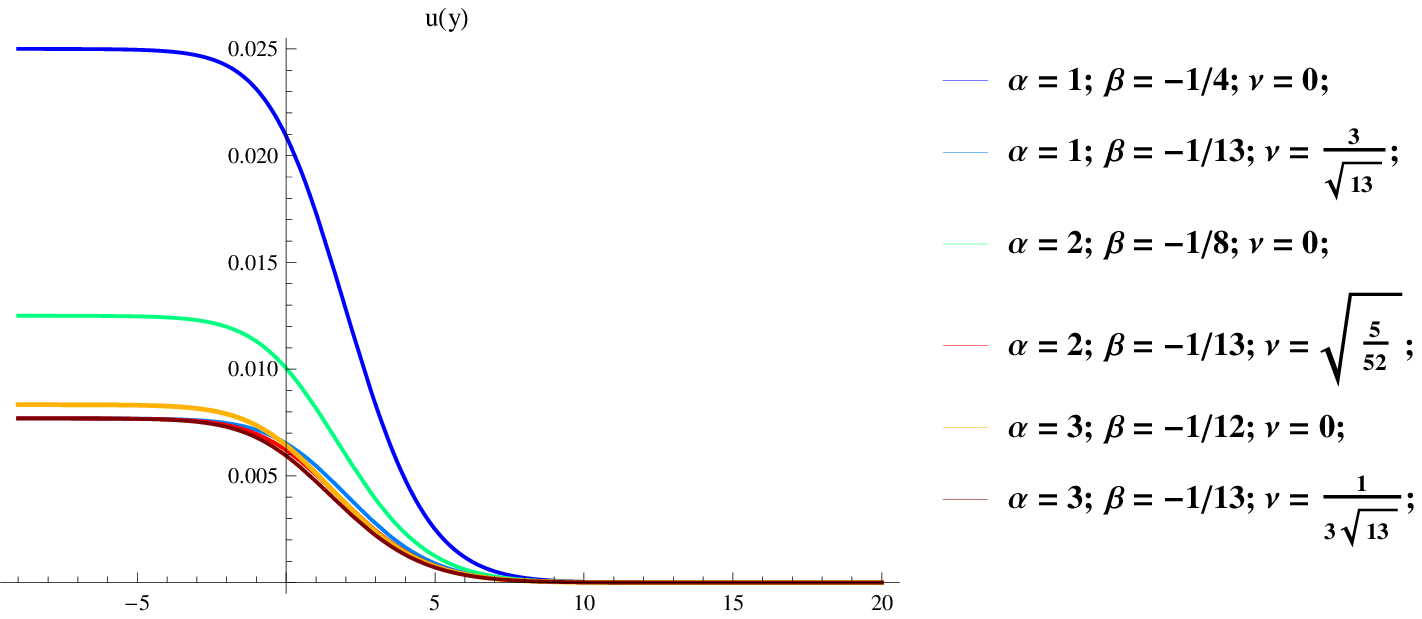} \\ Fig.4:  $ u(y) $; $c=1$; $C_{u}=9$; $ \sigma = 1 $; $ \textit{b}= 0.1 $}
\end{minipage}
\end{figure}

\subsubsection{$\textit{b}  < 0$}

Using the analysis of (10) one can see that the condition $ \textit{b} < 0$ along with $ \sigma = - 1 $ and $ C_{u} < 0 $ ($ \sigma C_{u} > 0 $) leads to the fact that the function $ u(y) $ has not changed, but $ v(y) $ becomes positive on all domain of definition. This function is presented in Fig.5. 

\begin{figure}[h!]
\center{\includegraphics[width=0.7\linewidth]{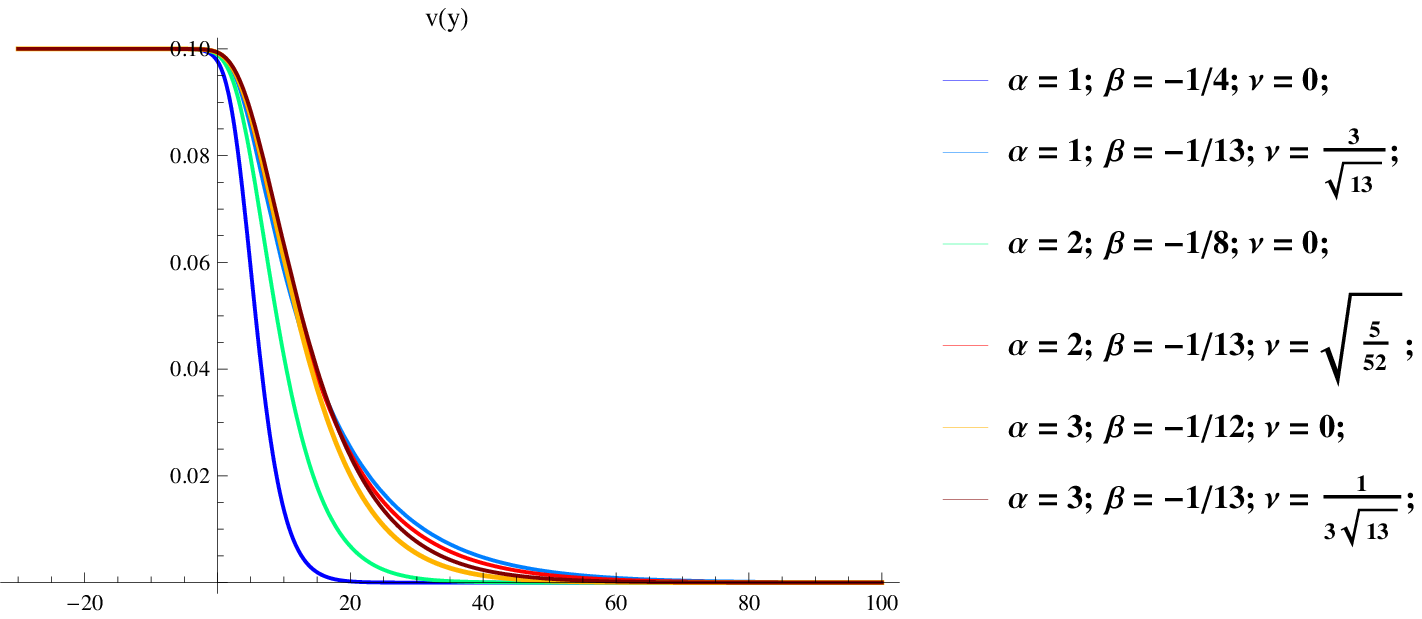} \\ Fig.5:  $ v(y) $; $c=1$; $C_{u}= - 9$; $ \sigma = - 1 $; $ \textit{b}= - 0.1 $}
\hfill
\end{figure}

\subsection{$\eta \neq 1$}

Let us return to equation (3) and rewrite it in terms of the variable $ \xi = e^{-\frac{cy}{\alpha}} $:
\be
\xi ^{2} \upsilon_{\xi \xi}  - \dfrac{\alpha \beta}{c^{2}} \,\upsilon + \dfrac{ \sigma \alpha C_{u}}{c^{2}} \, \xi ^{\alpha} \, \upsilon ^{\eta} = - \dfrac{\alpha \beta \textit{b}}{c^{2}}.
\ee
To integrate this equation we use the Lie group method of infinitesimal transformations \cite{Olver}. We find a group invariant of a second prolongation of one--parameter symmetry group vector of (11) and with its help we transform equation (11) into an equation of the first order. It turns out that nontrivial symmetry group requires some conditions:
\ba
\dfrac{\alpha \beta \textit{b}}{c^{2}} = 0,\\
\nonumber
\beta = \frac{(\alpha - 2)(\alpha + \eta + 1) c^{2}}{\alpha (\eta + 3)^{2}}
\ea
and we consider the case $ \textit{b} =0 $. Thus, $ \upsilon = v $ and for 
\ba
z & = & \dfrac{v^{\frac{1 - \eta}{\alpha}}}{y} \\
\nonumber
w & = & v_{y}\,\,v^{- \frac{\alpha + \eta - 1}{\alpha}}
\ea
we obtain the Abel equation of the second kind:
\be
w_{z}\,[(1 - \eta) w  -  \alpha z] + (\alpha + \eta - 1) z^{-1} w^{2} + \alpha z ( - \dfrac{\alpha \beta}{c^{2}} +  \dfrac{\sigma \alpha C_{u}}{c^{2}} z^{-\alpha}) = 0.
\ee
Then we find solutions of equation (14) in parametric form \cite{Z&P ODE} with the parameter $ t $. 
Now we consider the case $ 2 \alpha + \eta \neq  1$. A combination of substitutions leads to: 
\ba
z & = & \left( -\frac{(\eta + 3)[ (\eta +1)\,t^{2} + \frac{2\sigma \alpha C_{u}}{c^{2}} ]}{2(2\alpha + \eta -1)}\,\frac{\vartheta_{t}(t)}{\vartheta(t)}\right) ^{\frac{2}{\alpha}}\\
\nonumber
w & = & z^{\frac{2 - \alpha}{2}}\,\, \left( t + \frac{2(2\alpha + \eta +1)}{(\eta - 1)(\eta + 3)} z^{\frac{\alpha}{2}}  \right) \,\, + \frac{\alpha}{1 - \eta}\, z, 
\ea
where we  take 
\be
\vartheta(t) > 0 \,\,\, \text{and} \,\,\, (2\alpha + \eta -1) \,\vartheta_{t}(t) < 0, 
\ee
and equation (14) becomes an equation for the function $ \vartheta(t) $. Solving it, for $ \sigma C_{u} > 0 $ we obtain:
\be
\vartheta(t) = \tilde{C_{\vartheta}} \left( \dfrac{2 \sigma \alpha C_{u}}{c^{2}}\right) ^{-\frac{\eta + 3}{2(\eta + 1)}} \,t\,\,_{2}F_{1} ( \frac{1}{2}, \frac{\eta + 3}{2(\eta + 1)}; \frac{3}{2}; -\frac{(\eta + 1)c^{2}}{2 \sigma \alpha C_{u}} \, t^{2} ) + C_{\vartheta},
\ee
where $\tilde{C_{\vartheta}} $, $ C_{\vartheta} $ are constants and $ _{2}F_{1} $ is the hypergeometric Gauss function. Further we obtain the solutions of initial equations (2)--(3) in parametric form:
\ba 
y(t) & = & -\frac{\alpha (\eta + 3)}{c (2\alpha + \eta - 1)} \,\,\ln \bigg(   \vartheta(t)  \bigg) \\
\nonumber
v(t) & = & \left(-\dfrac{ \tilde{C_{\vartheta}} (\eta + 3)}{2 (2\alpha + \eta - 1)}\right)^{\frac{2}{1 - \eta}} \, \left( ( \eta + 1) t^{2} + \frac{2 \sigma \alpha C_{u}}{c^{2}} \right) ^{-\frac{1}{\eta + 1}} \, \bigg( \vartheta(t) \bigg)^{\frac{2 - \alpha}{2\alpha + \eta - 1}} \\
\nonumber
u(t) & = & C_{u} \left(-\dfrac{ \tilde{C_{\vartheta}} (\eta + 3)}{2 (2\alpha + \eta - 1)}\right)^{\frac{2}{1 - \eta}} \, \left( ( \eta + 1) t^{2} + \frac{2 \sigma \alpha C_{u}}{c^{2}} \right) ^{-\frac{1}{\eta + 1}} \, \bigg( \vartheta(t) \bigg)^{\frac{ \alpha \eta + 2\alpha + 2}{2\alpha + \eta - 1}} 
\ea
where the constant $ \tilde{C_{\vartheta}}  $ is chosen so that $ (2\alpha + \eta -1) \tilde{C_{\vartheta}}  < 0  $, what is consistent with (16). Using the asymptotic representation of hypergeometric Gauss function as $ t \rightarrow \pm  \infty $ \cite{Bateman&Erdelyi} we can take 
\be
C_{\vartheta} > | \tilde{C_{\vartheta}} | \,\, \frac{\pi}{2 \sqrt{\eta + 1} } \left( \frac{2 \sigma \alpha C_{u}}{c^{2}} \right) ^{-\frac{1}{\eta + 1}} \,\frac{\Gamma (\frac{1}{\eta + 1})}{\Gamma(\frac{\eta + 3}{2(\eta + 1)})}
\ee
in order for $ y, v $ and $ u $ be real. Then one can see that all functions (18) are continuous bounded ones for $ t \in \Re $ and $ v, u $ are positive. Hence, one may try biologically interpret the functions $ v(y) $ and $ u(y) $ and this requires further investigation. In Fig.6 one may see the different curves $ v(y) $ for $ \eta = 0.1 $ and different $ \alpha $. Fig.7 demonstrates $ v(y) $ and $ u(y) $ for two $ \eta < 1 $. Further, for larger values ​​of $ \alpha $ and $ \eta $ it seems more convenient to present curves $ y(t)$, $ v(t) $ and $ u(t) $ to analyze them, see Fig.8--Fig.10. One can see from (12) that $ \beta \gtrless 0 $ when $ \alpha \gtrless 2 $, and the case of $ \beta = 0 $, $ \alpha = 2 $ is presented in Fig.11. 
\begin{figure}[h!]
\begin{minipage}{0.49\linewidth}
\center{\includegraphics[width=1\linewidth]{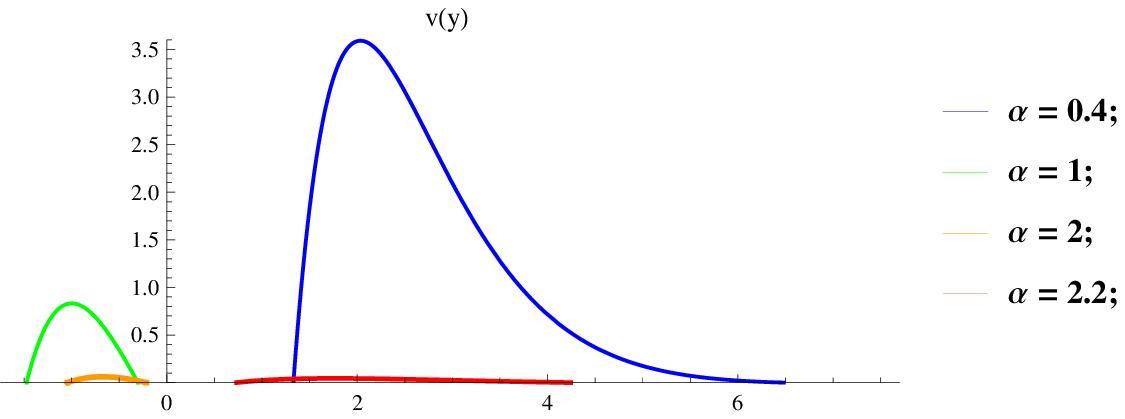} \\ Fig.6:  $ v(y) $;  $ \eta = 0.1 $; $ \frac{ \sigma \alpha C_{u}}{c^{2}} = 2 $; $c=1$; $C_{\vartheta} = 1.4 $; $ | \tilde{C_{\vartheta}} | =1 $;}
\end{minipage}
\hfill
\begin{minipage}{0.49\linewidth}
\center{\includegraphics[width=0.5\linewidth]{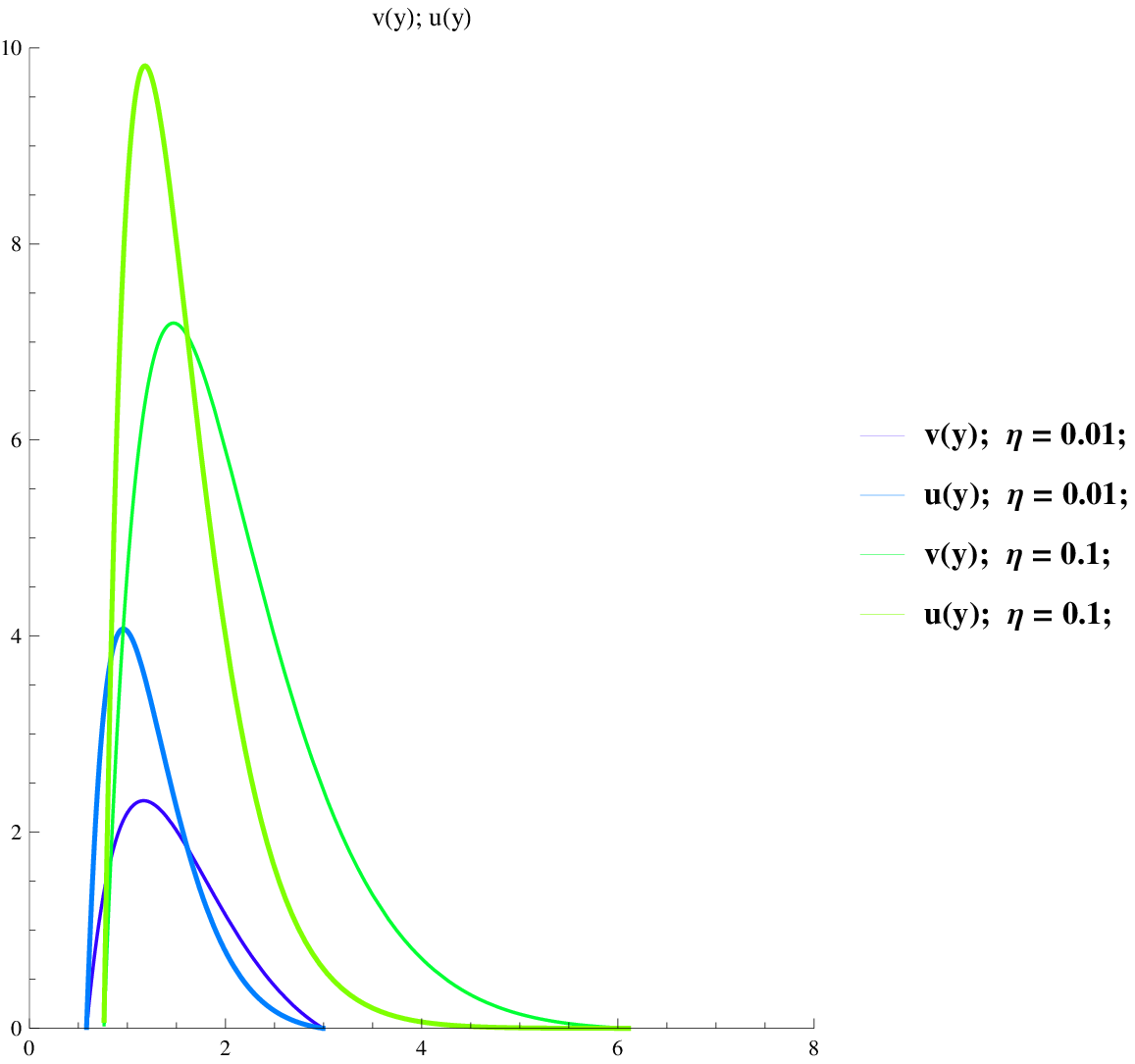} \\ Fig.7:  $ v(y) $; $ u(y) $; $ \alpha = 0.4 $; $ \frac{ \sigma \alpha C_{u}}{c^{2}} = 2 $; $c=1$; $C_{\vartheta} = 1.35 $; $ | \tilde{C_{\vartheta}} | =1 $;}
\end{minipage}
\end{figure}

\begin{figure}[h!]
\begin{minipage}{0.49\linewidth}
\center{\includegraphics[width=1\linewidth]{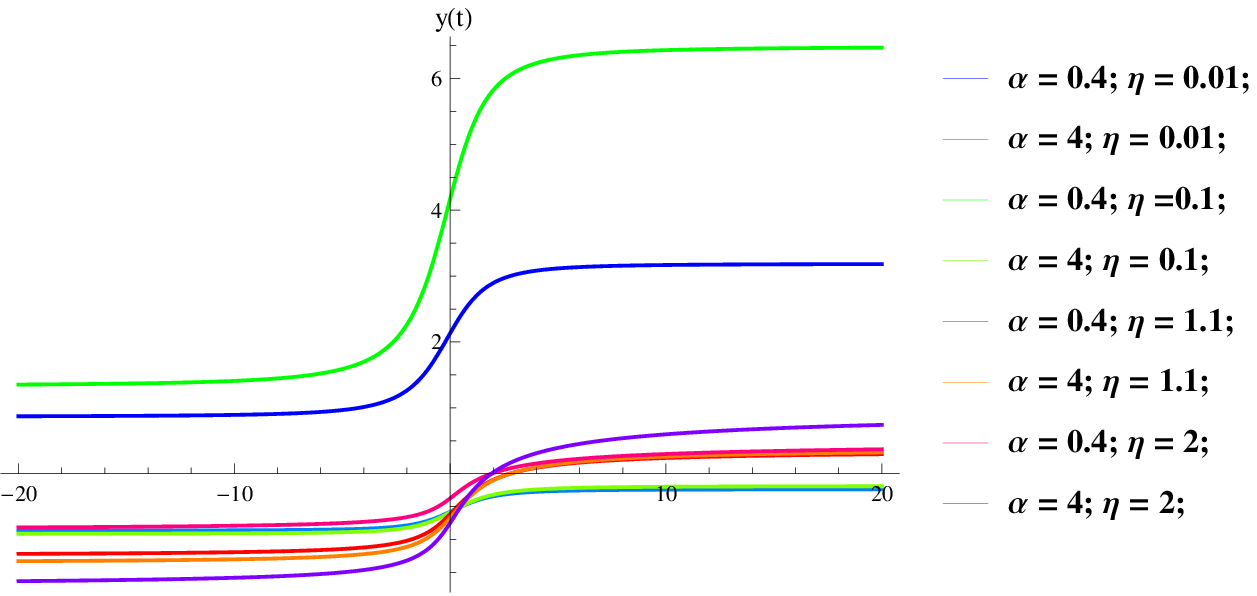} \\ Fig.8:  $ y(t) $; $ \frac{ \sigma \alpha C_{u}}{c^{2}} = 2 $; $c=1$; $C_{\vartheta} = 1.4 $; $ | \tilde{C_{\vartheta}} | =1 $; }
\end{minipage}
\hfill
\begin{minipage}{0.49\linewidth}
\center{\includegraphics[width=1\linewidth]{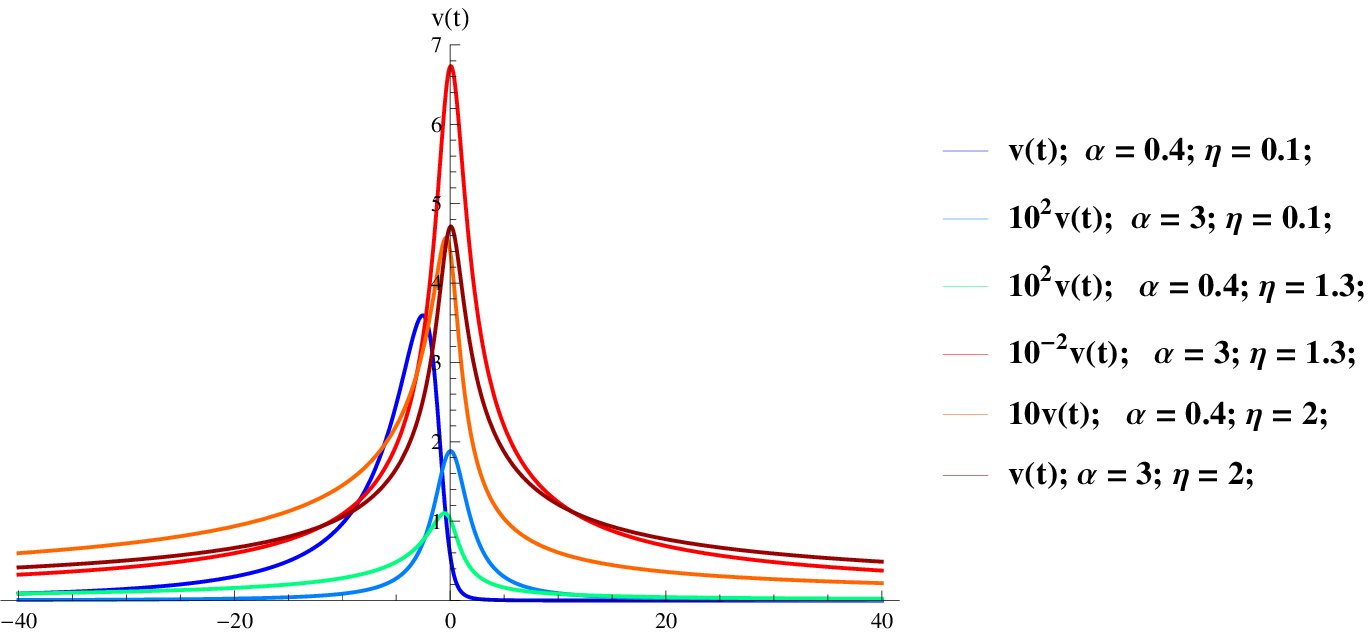} \\ Fig.9:  $ v(t) $; $ \frac{ \sigma \alpha C_{u}}{c^{2}} = 2 $; $c=1$; $C_{\vartheta} = 1.4 $; $ | \tilde{C_{\vartheta}} | =1 $;}
\end{minipage}
\hfill
\begin{minipage}{0.49\linewidth}
\center{\includegraphics[width=1\linewidth]{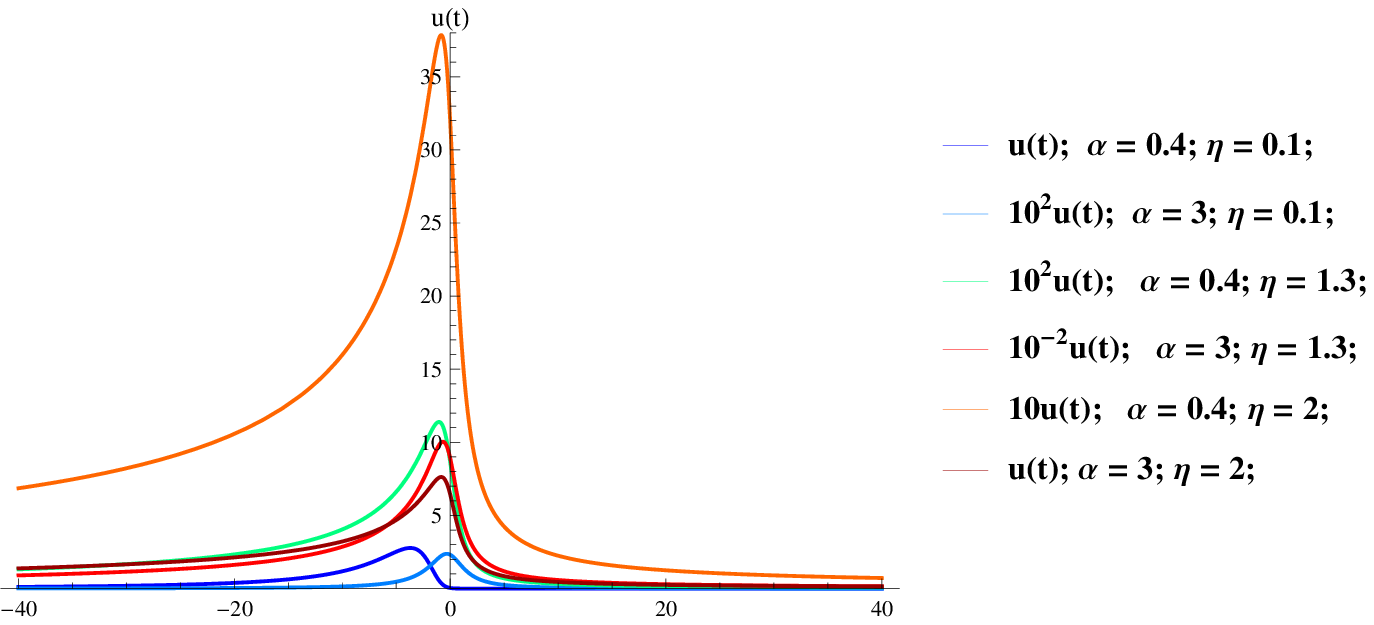} \\ Fig.10:  $ u(t) $; $ \frac{ \sigma \alpha C_{u}}{c^{2}} = 2 $; $c=1$; $C_{\vartheta} = 1.4 $; $ | \tilde{C_{\vartheta}} | =1 $;}
\end{minipage}
\hfill
\begin{minipage}{0.49\linewidth}
\center{\includegraphics[width=1\linewidth]{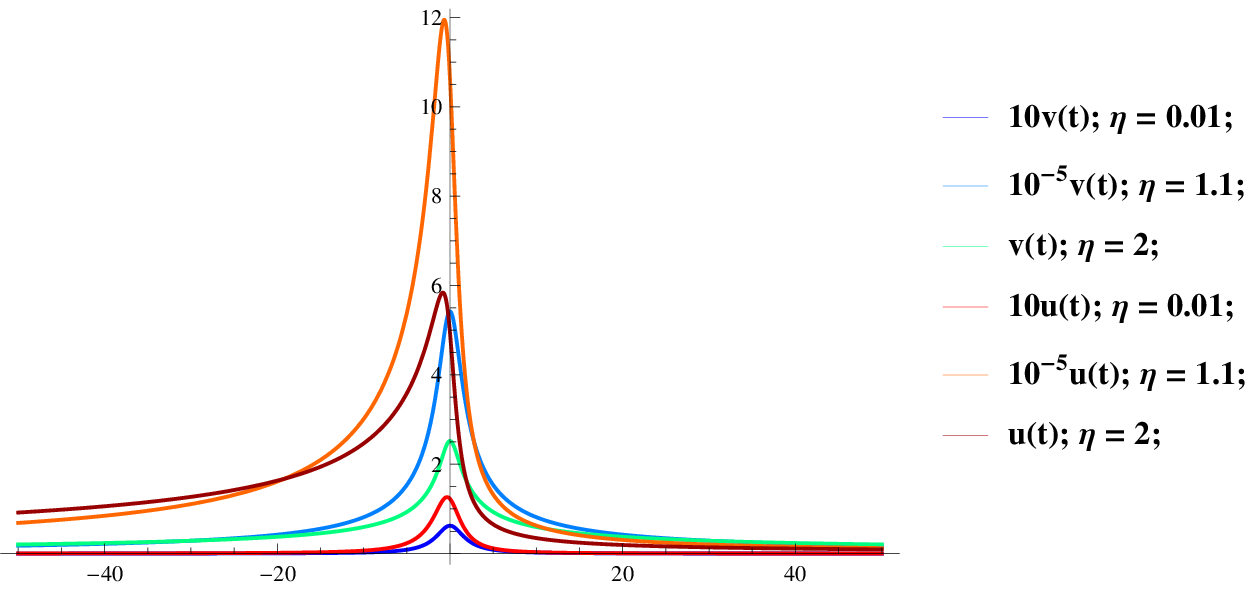} \\ Fig.11:  $ v(t) $; $ u(t) $; $ \alpha = 2 $; $ \frac{ \sigma \alpha C_{u}}{c^{2}} = 2 $; $c=1$; $C_{\vartheta} = 1.4 $; $ | \tilde{C_{\vartheta}} | =1 $;}
\end{minipage}
\end{figure}

\section{Logarithmic sensitivity}
\label{sec:2}

The model with logarithmic chemosensitivity function $ \phi (v) \sim  \ln v $ is also studied. For the case of $f(u, v) = - v^{m} u +  \tilde{\beta} v $, $ \tilde{\beta} = const $ an extensive analysis is performed in \cite{Wang}. This survey is focused on different aspects of traveling waves solutions. When $ m = 0 $ this model coincides with (1) for $ \textit{b} = 0 $. When $ \tilde{\beta} = 0 $ and $ m = 1 $ the system was studied in \cite{Nossal}, \cite{Rosen}. The complete analysis for $ \tilde{\beta} = 0 $ is performed in \cite{Wang}. An existence of global solution is established in \cite{W}. 

Now we consider the system with $ \phi (v) = \ln v $ and  $ f(u, v)= \tilde{\sigma} v u  - \tilde{\beta} v $. Similarly, a replacement $ t \rightarrow  \delta_{1} t $, $ u \rightarrow \sigma \dfrac{\tilde{\sigma} }{\delta_{1}} u $ gives $\delta_{1} = 1 $, $ \eta = \dfrac{\eta_{1}}{\delta_{1}} $, $ \alpha = \dfrac{\delta_{2}}{\delta_{1}} $, $ \beta = \dfrac{\tilde{\beta}}{\delta_{1}} $, $ \sigma = \pm 1  $. Then the model has the form:
\be 
\label{eq:20} 
\left\{
\begin{aligned}
u_{t}- u_{xx} + \eta (u \frac{v_{x}}{v})_{x} & =0 \\ 
v_{t}-\alpha v_{xx} - \sigma v u + \beta v  & =0.\\ 
\end{aligned}
\right.
\ee
Let us rewrite system (20) in terms of function $ \upsilon (x, t) = \ln v(x, t) $:
\be 
\left\{
\begin{aligned}
u_{t}- u_{xx} + \eta (u \upsilon_{x})_{x} & = 0 \\ 
\upsilon_{t}- \alpha \upsilon_{xx} - \alpha  (\upsilon_{x})^{2} + \beta  - \sigma u & = 0,\\ 
\end{aligned}
\right. 
\tag{\ref{eq:20}$'$}
\ee
then in terms of traveling wave variable $ y = x - ct $, $ c = const $, (20$'$) has the form:
\be \left\{
\begin{aligned}
u_{y} + c u -  \eta u \upsilon_{y}+ \lambda & =0 \\ 
\alpha \upsilon_{yy} + \alpha (\upsilon_{y})^{2} + c \upsilon_{y} - \beta + \sigma u & =0,\\
\end{aligned}
\right. 
\tag{\ref{eq:20}$'*$} 
\ee
where $ u=u(y) $, $ \upsilon = \upsilon (y) $ and $ \lambda $ is an integration constant. To integrate (20$'*$) we tested this system on the Painlev\'e ODE test. One can show that for $ \eta > 0 $ it passes this test only if $ \alpha = 2 $ with the additional condition $ \lambda = - \sigma c \beta \left( 1 + \dfrac{\eta}{2} \right) $ \cite{MSh_ArXiv}. If we express $ u(y) $ as $ \upsilon(y) $ from (20$'*$), we obtain an equation only for $ \upsilon (y) $; for $ \alpha = 2 $ it has the form:
\be
2\upsilon_{yyy} + 3c \upsilon _{yy} + (c^{2} + \eta \beta ) \upsilon_{y} + 2(2 - \eta )\upsilon_{y} \upsilon_{yy} + 2(2 - \eta ) (\upsilon_{y})^{2} -2\eta (\upsilon_{y})^{3} - c\beta - \sigma \lambda = 0.
\ee
For $ \lambda = - \sigma c \beta \left( 1 + \dfrac{\eta}{2} \right) $ this equation can be linearized. It becomes equivalent to the following linear equation for $ F $:
\be
F_{y} + c F = 0, \,\,\,\text{where} \,\, \,\,\,    
F(y) = e^{2 \upsilon} \left( 2 \upsilon_{yy} + c \upsilon_{y} - \eta (\upsilon_{y})^{2} + \dfrac{\eta \beta} {2} \right)  
\ee
that gives the equation for $ \upsilon (y) $:
\be
2 \upsilon_{yy} + c \upsilon_{y} - \eta (\upsilon_{y})^{2} + \dfrac{\eta \beta} {2} = C_{F} e^{-2 \upsilon -cy},
\ee
$ C_{F} = const $. If we rewrite (23) in terms of the variable $ \xi = e^{-\frac{cy}{2}} $ for the function $ \Psi (\xi) = e^{- \frac{\eta}{2} \upsilon} $ we obtain an equation similar to (11) with zero right-hand side:
\be
\xi ^{2} \Psi_{\xi \xi}  - \dfrac{\eta^{2} \beta}{2c^{2}} \,\Psi + \dfrac{\eta C_{F}}{c^{2}} \, \xi ^{2} \, \Psi ^{\frac{4}{\eta} + 1} = 0.
\ee
Using the result of the symmetry group analysis of (11) we can write solution for $ \beta = 0 $ (see (18)):
\ba 
y(t) & = & -\frac{2 }{c} \,\,\ln \bigg(   \vartheta(t)  \bigg) \\
\nonumber
v(t) & = & \dfrac{ |\tilde{C_{\vartheta}}|}{2 } \, \left( 
\frac{2( \eta + 2)}{\eta} \,\, t^{2} + \frac{2 \eta C_{F}}{c^{2}} \right) ^{\frac{1}{\eta + 2}} \\
\nonumber
\ea
where $ \vartheta (t) $ is given in (17) and $ u(y) $ may be expressed from (20$'*$). However one may see that $ v \rightarrow \infty $ as $ t \rightarrow \pm \infty $ and this solution is unacceptable as a biological function.

Another possibility to solve this equation exactly is to put $ C_{F}$ equal to zero. When $ C_{F} = 0 $, that means $ F(y) = 0 $, and $ \beta \neq 0 $ equation (24) can be linearized by $ \xi = e^{\tau} $ \cite{Z&P ODE}. Its solution has three forms according to a sign of the expression $ D = \dfrac{2 \eta^{2} \beta}{c^{2}} + 1$. Since $v$ should be nonnegative and bounded function as $ cy \rightarrow \pm \infty $ the only suitable solution is
\ba
v(y) & = & e^{\frac{c}{2 \eta}\, y} \,\, \left( C_{-}\,e^{-\frac{c \sqrt{D} }{4}\, y}  + C_{+}\, e^{\frac{c \sqrt{D}}{4}\, y}    \right) ^{-\frac{2}{\eta}}
\ea
where $ C_{\pm} $ are positive constants and $ \beta > 0 $. Unfortunately, the corresponding solution for $ u(y) $ is alternating and has the form:
\ba
u(y) & = & -\frac{\sigma c^{2}\, (\eta + 2)} {2 \eta ^{2}}\, ( C_{-}^{2} (1 + \sqrt{D}) e^{-\frac{c \sqrt{D} }{4}\, y} +  C_{+}^{2} (1 - \sqrt{D}) e^{\frac{c \sqrt{D} }{4}\, y} \\
\nonumber
& - & \frac{4 \eta^{2 } \beta }{c^{2}} C_{-} C_{+} )  \, \left( C_{-}\,e^{-\frac{c \sqrt{D} }{4}\, y}  + C_{+}\, e^{\frac{c \sqrt{D}}{4}\, y}    \right) ^{-\frac{2}{\eta}}
\ea
It is easy to see what $ \sigma u(y) \rightarrow \frac{c^{2}(\eta + 2)} {2 \eta ^{2}} (-1 \pm \sqrt{D}) $ as $ cy \rightarrow \pm \infty $. These functions are presented in Fig.12--Fig.13.
\begin{figure}[h!]
\begin{minipage}{0.49\linewidth}
\center{\includegraphics[width=0.9\linewidth]{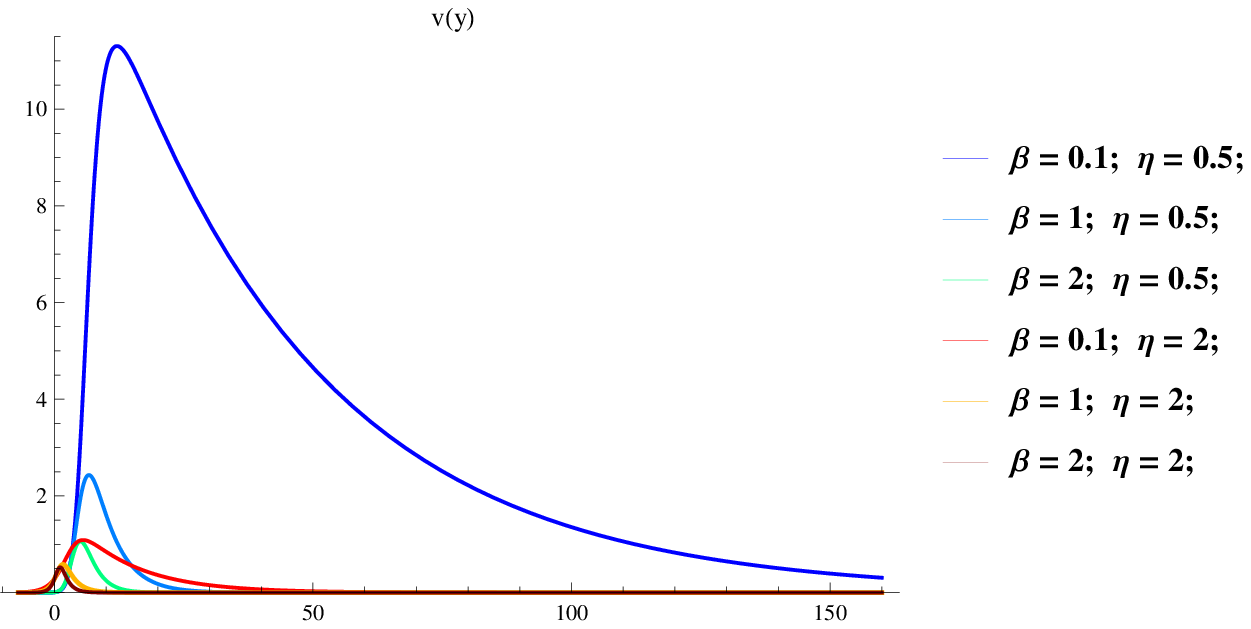} \\ Fig.12:  $ v(y) $; $c=1$;}
\end{minipage}
\hfill
\begin{minipage}{0.49\linewidth}
\center{\includegraphics[width=0.9\linewidth]{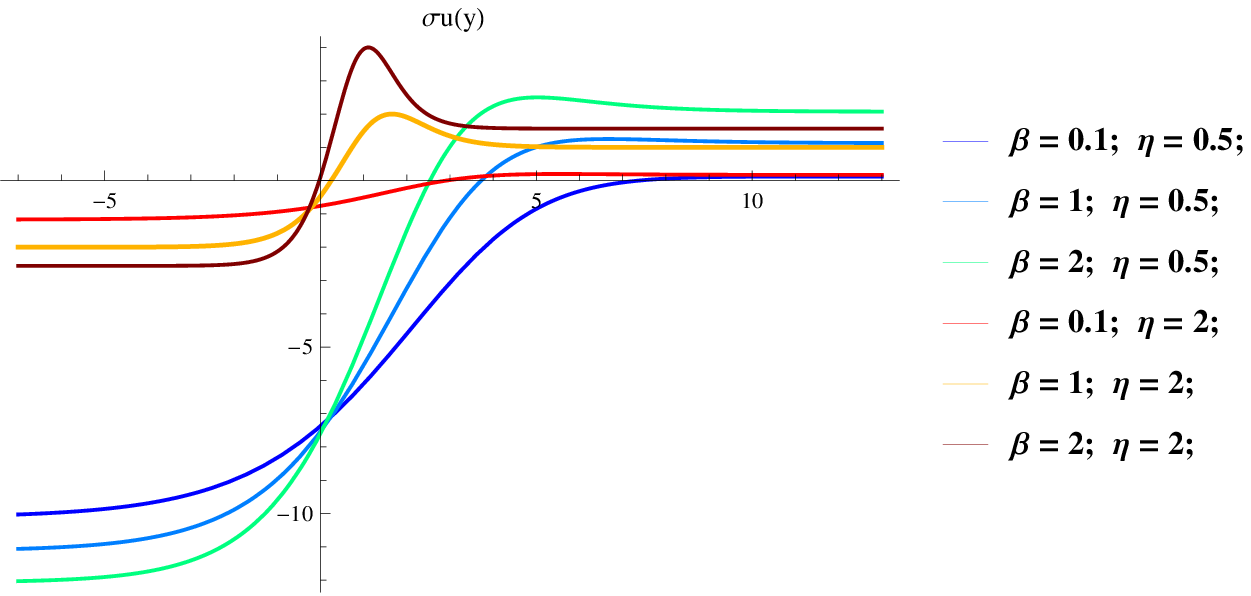} \\ Fig.13:  $ \sigma u(y) $; $c=1$;}
\end{minipage}
\end{figure}

\section{Linear sensitivity}
\label{sec:3}

Let us consider the system with linear function $ \phi (v) \sim v $. When $ f(u, v) = u - v $ the system is called the minimal chemotaxis model following the nomenclature of \cite{Childress&Percus}. This model is often considered with $ f(u, v) = \tilde{\sigma} u - \tilde{\beta} v $ ($ \tilde{\sigma} $ and $ \tilde{\beta}  $ are constants) and it is studied in many papers. As was proved in \cite{Osaki&Yagi}, \cite{Hillen&Potapov} the solutions of this system are global and bounded in time for one space dimension. The case of positive $  \tilde{\sigma} $ and nonnegative $ \tilde{\beta} $ is studied in \cite{J&L}-\cite{F}. As we noted earlier, a sign of $ \tilde{\sigma} $ may effect on mathematical properties of the system, what changes its solvability conditions \cite{TF}. The review article \cite{H1} summarizes different mathematical results. 

Now we consider the linear chemosensitivity function $ \phi (v) = v $ and $ f(u, v)= \tilde{\sigma} u - \tilde{\beta} v  $. The replacement $ t \rightarrow  \delta_{1} t$, $ v \rightarrow \dfrac{\eta_{1}}{\delta_{1}} v $, $ u \rightarrow \sigma \dfrac{\tilde{\sigma} \eta_{1}}{\delta_{1}^{2}} u $ leads to $\delta_{1} = \eta_{1} = 1 $, $ \alpha = \dfrac{\delta_{2}}{\delta_{1}} $, $ \beta = \dfrac{\tilde{\beta}}{\delta_{1}} $, $ \sigma = \pm 1  $. Then the system has the form: 
\be 
\label{eq:123} 
\left\{
\begin{aligned}
u_{t}- u_{xx} + (u v_{x})_{x} & =0 \\ 
v_{t}-\alpha v_{xx} + \beta v - \sigma u & =0.\\ 
\end{aligned}
\right.
\ee

This system reduces to system of ODEs in terms of traveling wave variable $ y = x - ct $, $ c = const $:
\be 
\left\{
\begin{aligned}
u_{y}+cu-u v_{y}+ \lambda & =0 \\ 
\alpha v_{yy}+cv_{y} - \beta v + \sigma u & =0,\\
\end{aligned}
\right.
\tag{\ref{eq:123}$*$} 
\ee
where $ u=u(y) $, $ v=v(y) $ and $ \lambda $ is an integration constant. As shown in \cite{MSh} this system passes the Painlev\'e ODE test only if $ \alpha = 2 $ and $ \beta = 0 $. Consequently, in this case we can solve ($28*$) and the exact solution has the form \cite{MSh}:
\ba
v & = & - \ln \left[  e^{-\frac{cy}{2}}\,\,A^{2}\,\left( I_{\nu}(\frac{\kappa}{|c|}\,\,e^{-\frac{cy}{2}})\,+\,B\,K_{\nu}(\frac{\kappa}{|c|}\,\,e^{-\frac{cy}{2}}) \right)^{2} \right] \\ \nonumber
 u & = & - \sigma \left( (v_{y})^{2} -  \kappa^{2}\,e^{-cy} + \dfrac{\lambda}{c} \right) , \,\,\,\, \text{where $ \nu^{2}=\dfrac{1}{4}-\dfrac{\lambda}{c^{3}} $}, 
\ea
$  \kappa >0 $, $ A $ and $ B $ are arbitrary constants. The functions $ I_{\nu} $ and $ K_{\nu} $ are Infeld's and Macdonald's functions respectively (Bessel's functions of imaginary argument). This solution is not satisfactory from the biological point of view, since $ v(y) $ is an alternating function for any $ \nu $. However it seems interesting because of the following: in the case of $ \nu =  \dfrac{1}{2} $ and $ B = \frac{2 + \pi}{2 \pi} $ in terms of $ e^{-\frac{cy}{2}} $ its form coincides with the well-known Korteweg-de Vries soliton
\ba
e^{v(\,e^{-\frac{cy}{2}})} = \frac{\kappa}{C^{2}|c|}\, sech^{2}\left( \frac{\kappa}{|c|}\,\,e^{-\frac{cy}{2}} + \dfrac{1}{2} \ln \dfrac{2}{\pi}\right). 
\ea
For  $ \nu = \dfrac{1}{2} $ and arbitrary $ B $ the function $ u(y) $ is
\be
u(y)  = \dfrac{ \sigma (\pi \, B - 1 )\,\kappa^{2}\,e^{-cy} }{\left( \sinh (\dfrac {\kappa}{|c|}\,\,e^{-\frac{cy}{2}})  +  \dfrac{\pi}{2} \, B\, e^{- \dfrac{\kappa}{|c|}\,\,e^{-\frac{cy}{2}}}  \right)^{2}  }. 
\ee
One can see that for $ \sigma = 1 $ (an increase of a chemical substance) the cells density $ u(y) \geq 0 $ for $ B \geq \dfrac{1}{\pi} $, and that for $ B > 0 $ $ u(y) $ is the solitary continuous solution vanishing as $ y \rightarrow \pm \infty $, whereas for $ B < 0 $ $ u(y) $ has a point of discontinuity. One can say that when $ B < 0 $ we obtain "blow up" solution in the sense that it goes to infinity for finite $ y $, and this is true for different $ \nu $. The functions (29) for  $ \nu = \dfrac{1}{2} $ are presented in Fig.14--Fig.15.
\begin{figure}[h!]
\begin{minipage}{0.49\linewidth}
\center{\includegraphics[width=0.9\linewidth]{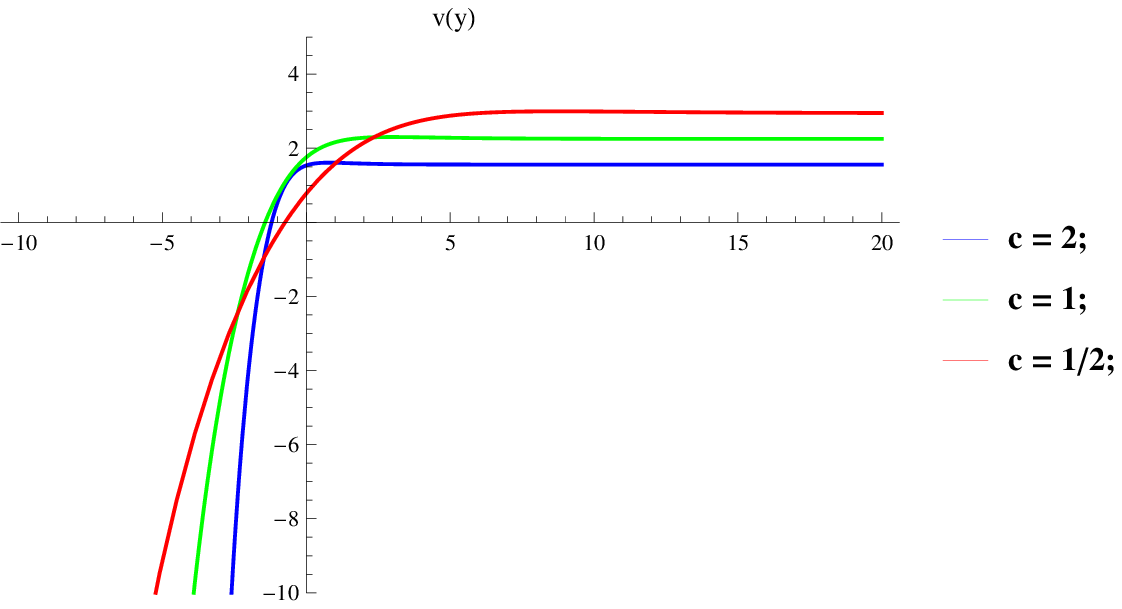} \\ Fig.14:  $ v(y) $; $k=1$; $ B = \frac{2 + \pi}{2 \pi} $;}
\end{minipage}
\hfill
\begin{minipage}{0.49\linewidth}
\center{\includegraphics[width=1\linewidth]{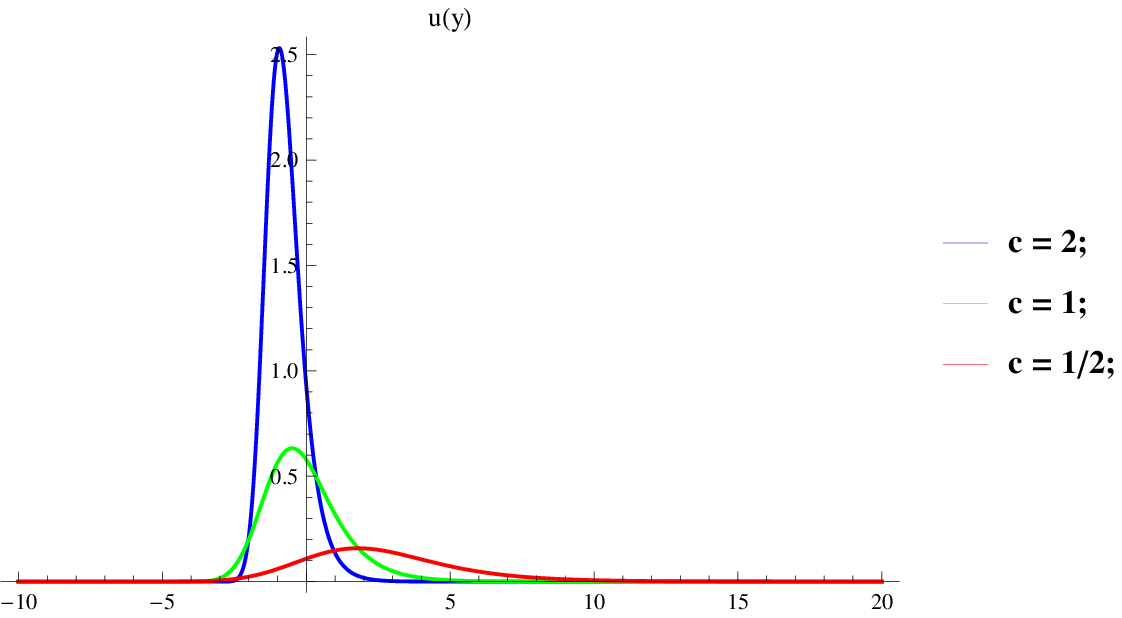} \\ Fig.15:  $ u(y) $; $ \sigma = 1 $; $k=1$; $ B = \frac{2 + \pi}{2 \pi} $;}
\end{minipage}
\end{figure}

\section{Conclusion}
\label{sec:5}

We investigate three different one-dimensional parabolic-parabolic Patlak-Keller-Segel models. For each of them we obtain the exact solutions in terms of traveling wave variables. Not all of these solutions are acceptable for biological interpretation, but there are solutions that require detailed analysis. It seems interesting to consider the latter for the experimental values ​​of the parameters and see their correspondence with experiment. This question requires a further investigations.

\end{document}